# Probing Early Clustering with Ly$\alpha$ Absorption Lines Beyond the Quasar Redshift

Abraham Loeb and Daniel J. Eisenstein[1]

Astronomy Department, Harvard University,
60 Garden St., Cambridge MA 02138

## ABSTRACT

Groups and clusters of galaxies hosting a quasar can be found through the detection of Ly$\alpha$ absorption lines beyond the quasar redshift. The effect occurs whenever the distortion to the redshift distribution of Ly$\alpha$ clouds induced by the cluster potential extends beyond the proximity effect of the quasar. Based on CDM cosmological models for the evolution of structure, we predict the probability for finding lines beyond the quasar redshift ($z_{\rm abs} > z_Q$) under the assumption that the physical properties of Ly$\alpha$ clouds are not affected by flows on large scales ($\gtrsim$ Mpc) in the quasi-linear regime. If quasars randomly sample the underlying galaxy distribution, the expected number of lines with $z_{\rm abs} > z_Q$ per quasar can be as high as $\sim 0.5 \times [(dN/dz)/350]$ at $z = 2$, where $dN/dz$ is the number of Ly$\alpha$ lines per unit redshift far from the quasar. The probability is enhanced if quasars typically reside in small groups of galaxies. A statistical excess of Ly$\alpha$ lines is expected near very dim quasars or around metal absorption systems. Due to clustering, the standard approach to the proximity effect overestimates the ionizing background flux at high redshifts by up to a factor of $\sim 3$. This result weakens the discrepancy between the deduced background flux and the contribution from known populations of quasars.

*Subject headings:* cosmology: theory–quasars: general

---

[1]Also at: Physics Department, Harvard University

## 1. Introduction

Acceptable cosmological models for structure formation are designed to agree with large-scale-structure observations in the local universe. Despite the common calibration to fit the universe at present, these models can have very different predictions about its past history (e.g. Mo & Miralda-Escudè 1994). In this paper, we discuss a method to probe clustering of galaxies at high redshifts ($z \sim 1$–5) that could potentially test such models. The method makes use of the redshift distribution of Ly$\alpha$ clouds in the vicinity of a quasar.

If a quasar resides in a cluster or group of galaxies, its peculiar velocity may shift its apparent redshift below the redshift of its nearest Ly$\alpha$ cloud. This effect is further enhanced by the gravitational potential of the massive host, which induces infall of distant Ly$\alpha$ clouds and increases the number density of clouds out to scales of ∼5 Mpc around the cluster. On the other hand, the quasar reduces the number of absorbers by ionizing clouds that are in physical proximity to it (Weymann, Carswell & Smith 1981; Carswell et al. 1982; Murdoch et al. 1986; Tytler 1987; Bajtlik, Duncan, & Ostriker 1988, hereafter BDO; Kovner & Rees 1989; Lu, Wolfe, & Turnshek 1991, hereafter LWT; Bechtold 1994). Nevertheless, as we show later, the redshift distortion due to clusters can extend beyond the effective radius of the quasar proximity effect.

The presence of an excess of C IV absorption systems near or beyond the emission line redshift of quasars has been discussed for two decades (see Foltz et al. 1987 for a review). The observed excess was attributed to absorption within clusters of galaxies or within the quasar host galaxies (Bahcall 1975; Weymann & Williams 1978). In the near future, spectroscopic data from the Keck telescope (Sargent 1994) and the Sloan Digital Sky Survey (Gunn & Knapp 1993) are expected to shed more light on the statistics of this phenomenon. In this work we use the current theory for structure formation in the universe in order to quantify the statistical effect that clusters should have on the redshift distribution of absorption systems (Sargent et al. 1980; Young et al. 1982). We prefer to focus on Ly$\alpha$ rather than C IV absorption systems since Ly$\alpha$ clouds are much more common and therefore sample better the cosmological mass distribution around clusters. In addition, the proximity effect of the quasar eliminates the contribution of clouds that physically reside inside the host cluster. This simplifies the theoretical analysis since it leaves out the unknown properties of absorbers inside virialized systems such as galaxies or clusters. The clustering signature is then dominated by the peculiar velocity shear on the scale of a few Mpc where the matter has not virialized yet. We consider the statistics of lines with arbitrary velocity shifts relative to the quasar out to $-2000\,\mathrm{km/s}$.

The outline of this paper is as follows. In §2 we calculate the probability for finding

Lyα lines above the quasar redshift. The calculation is done under the assumptions that quasars are drawn at random from the underlying distribution of either galaxies ($\sim 10^{12} M_\odot$) or small groups of galaxies ($\sim 10^{13} M_\odot$), and that the physical properties of Lyα clouds are not altered by cosmological flows on large scales ($\gtrsim$ Mpc) in the linear and quasi-linear regimes. The clustering signature obviously depends on the assumed minimum mass of the quasar host and on the threshold for cloud destruction; we therefore present results for different sets of assumptions. The current distribution of cluster properties (Bahcall & Cen 1993; Zabludoff et al. 1993) is evolved back in time using the Press-Schechter formalism (Press & Schechter 1974) and the spherical collapse model (Gunn & Gott 1972); a procedure known to be in good agreement with N-body simulations (Lacey & Cole 1993 and references therein). Since the mass scales of interest are covered by existing galaxy surveys, we adopt a power spectrum of initial density perturbations that gives a good phenomenological fit to local cluster data (cf. Bahcall & Cen 1993). Numerical results from this approach are presented in §3. Finally, §4 summarizes our main conclusions.

## 2. Method

Previous analyses of the redshift distribution of Lyα clouds in the vicinity of a quasar (e.g. BDO, LWT, Bechtold 1994) have assumed that the velocity difference between the absorbing Lyα cloud and the quasar is a measure of the actual physical distance between the two. In this paper we include two sources of peculiar velocity that alter this assumption. First, we allow the quasar to reside in a virialized group of galaxies and to have a Gaussian probability distribution of peculiar velocities according to the velocity dispersion of this host. Second, we allow for an infall peculiar velocity and for a radial displacement of the Lyα clouds far outside the host due to its gravitational pull. This infall also increases the number density of Lyα clouds near the quasar. We treat the distribution of clustering properties at high redshifts according to the Press-Schechter formalism and approximate the infall dynamics by the spherical collapse model.

We begin with a brief review of the spherical collapse solutions, as they play an important role in our analysis. In a bound system, the solution $r(t)$ for the radius of a spherical shell with a mass $M$ interior to it may be written as,

$$\begin{aligned} r &= A(1 - \cos \eta) \\ t &= B(\eta - \sin \eta), \end{aligned} \quad (1)$$

where $A^3 = GMB^2$ (Gunn & Gott 1972; Peebles 1980). The remaining degree of freedom in the choice of $A$ and $B$ may be set by picking the mean overdensity interior to the shell $\bar{\delta}(R) = \int_{r<R} d^3x\, \delta(\mathbf{x})$ at a particular redshift. Here $\delta(\mathbf{x}) = [\delta\rho(\mathbf{x}) - \rho_b]/\rho_b$, $\rho_b$ is the background density, and $R$ is the radius of the shell. If the overdensity is $\bar{\delta}_i$ at a high redshift $z_i$ ($\eta_i \ll 1$), then

$$B = \frac{1}{2H_0\sqrt{\Omega_0}}\left(\frac{5}{3}\bar{\delta}_i(1+z_i) - (\Omega_0^{-1} - 1)\right)^{-3/2}, \qquad (2)$$

where $H_0$ and $\Omega_0$ are the Hubble constant and density parameter of the background universe today. Note that this solution has the peculiar velocities of the growing mode. For an unbound system, we alter equation (1) to the usual hyberbolic solution and change equation (2) to

$$B = \frac{1}{2H_0\sqrt{\Omega_0}}\left(\Omega_0^{-1} - 1 - \frac{5}{3}\bar{\delta}_i(1+z_i)\right)^{-3/2}. \qquad (3)$$

Using the growing mode growth factor, one can calculate the *r.m.s.* fluctuations $\sigma_M(z_i)$ on a particular mass scale $M$ at redshift $z_i$. This yields $\sigma_M(z_i)(1+z_i) = \sigma_M(0)$ for an $\Omega_0 = 1$ universe and $\sigma_M(z_i)(1+z_i) = \sigma_M(0)D_1(\Omega_0^{-1}-1)/[2(\Omega_0^{-1}-1)/5]$ for an $\Omega_0 < 1$ universe, where $D_1(x) = 1 + 3/x + (3\sqrt{1+x}/x^{3/2})\ln(\sqrt{1+x} - \sqrt{x})$ is the growth factor in an open universe (Groth & Peebles 1975). In this manner, one can connect the Gaussian initial conditions to the properties of the spherical solution.

Given a particular background cosmology and a power spectrum of Gaussian initial perturbations, the Press-Schechter formalism and the spherical collapse model yield the distribution of object masses as a function of time. We assume that an object has virialized when its outermost shell reaches half its turn-around radius (i.e. $\eta = 3\pi/2$). For a given cosmology, there is one value of $\bar{\delta}_i(1+z_i)$ that results in a collapse at the epoch of the quasar's emission $t_Q$. The ratio of this value and $\sigma_M(z_i)(1+z_i)$ yields the critical peak height threshold $\nu_c(t_Q, M)$ needed to get a collapse on a given mass scale by the time $t_Q$. The Press-Schechter model then predicts the number density of collapsed objects $n$ at time $t_Q$,

$$\frac{dn}{d\ln M} = \sqrt{\frac{2}{\pi}}\left(\frac{d\ln\nu_c}{d\ln M}\right)\frac{\rho_b}{M}\nu_c(t_Q, M)e^{-\nu_c^2/2}. \qquad (4)$$

We set the velocity dispersion of a virialized object as follows. The specific binding energy of the system at turn-around is roughly $-GM/r_{\rm ta}$, where $M$ is the mass of the system. The turn-around radius $r_{\rm ta}$ is found by setting the collapse time of the object ($\eta = 3\pi/2$) to $t_Q$. Assuming that the system relaxes into an isothermal sphere with a one-dimensional velocity dispersion $\sigma$, the specific kinetic energy of the final virialized object is $\frac{3}{2}\sigma^2$. The virial theorem implies that this kinetic energy is equal to the absolute value of the total energy, yielding

$$\sigma = \left[\left(\frac{3\pi}{2}+1\right)\frac{GM}{3t_Q}\right]^{1/3}. \qquad (5)$$

One should note that estimates of this type appear with a variety of proportionality constants in the literature (Narayan & White 1987; White et al. 1993), but the exact number depends on the true density profile and collapse history of the object.

For each object, we assume that the probability of hosting a quasar is proportional to its mass, but in addition we assume that there is a minimum mass for a quasar host, chosen to be either $10^{12}\,M_\odot$ or $10^{13}\,M_\odot$ (all masses refer to the epoch of the quasar's emission). The former choice corresponds to galactic masses, which appear necessary to power the quasar (Turner 1991), while the latter choice corresponds to masses of small galaxy groups, which are suggested by the clustering properties of quasars (Bahcall & Chokshi 1991). For a given quasar, the probability distribution of host masses is proportional to $M\,dn/dM$. By determining the host mass, we also find the velocity dispersion of the host, allowing us to randomly select the line-of-sight peculiar velocity of the quasar from a Gaussian distribution. This procedure is clearly appropriate for groups of galaxies, but it is inappropriate if the host is just a single galaxy with the massive quasar black hole at its center. We ignore the latter case because $10^{12}\,M_\odot$ host systems could still be composed of unmerged components at high redshifts and because the velocity dispersions of single galaxies are quite low and make a small contribution to the clustering signature.

Let us now turn to the peculiar velocities of the Ly$\alpha$ clouds. Far away from the quasar, the clouds are uniformly distributed with no significant correlations (Sargent et al. 1980, Webb & Barcons 1991). We therefore assume that prior to the application of the quasar's ionizing flux, the clouds trace the mass distribution around the cluster. This means that before gravitational interactions cause the region around the host to contract, the clouds are uniformly distributed. Previous studies of quasar spectra have found that far from the quasar the number of Ly$\alpha$ lines per unit redshift is well described by a power-law,

$$dN/dz = A_0(1+z)^\gamma, \qquad (6)$$

with $\gamma = 2.4\pm0.3$ and $A_0 \approx 3$ for a rest equivalent width $W > 0.36$Å (BDO, LWT), although more recently Bechtold (1994) found a shallower redshift dependence with $\gamma = 1.9 \pm 0.3$ for $W > 0.32$Å and $\gamma = 1.3 \pm 0.2$ for $W > 0.16$Å. The values of $A_0$ and $\gamma$ depend on the sensitivity and the spectral resolution of the observations. In this work, we use $dN/dz$ values of 350 at $z = 2$ and 700 at $z = 4$ as rough estimates for the expected density of lines in forthcoming observations with the Keck telescope (Sargent 1994). According to equation (6), these values correspond roughly to a rest equivalent-width threshold of $\sim 0.02$Å or a column density threshold of $\sim 4 \times 10^{12}$ cm$^{-2}$. Our results are linearly proportional to these choices for $dN/dz$. We then find the number of lines per unit distance using the relation $dN/dr = (H_Q/c)(1+z_Q)(dN/dz)$, where $H_Q$ is the Hubble constant at the redshift of the quasar and $c$ is the speed of light.

To evaluate the gravitational effect of the host cluster on the Ly$\alpha$ cloud distribution, we consider the quasar host and its environment at very early times and fix the origin of the coordinate system at the center of the host. The host material comes from a sphere centered at the origin with a mass equal to the final host mass. We assume that no Ly$\alpha$ clouds ultimately survive inside the host because of its hot virialized gas and the intense ionizing radiation of the quasar. Thus, the material that eventually makes up the Ly$\alpha$ clouds originates outside this sphere. Starting from the outer radius of the host sphere $r_{\rm host}$, the clouds are assumed to be uniformly distributed at their initial positions and to behave as test particles. We then evolve a particular cloud position forward in time using the spherical solution. In doing so, we need only specify the mean overdensity interior to the radius of the cloud. The spherical solution yields the position and velocity at the final time (i.e. the time of observation). The final cloud velocity, which includes both the Hubble flow and the peculiar velocity, is added to the quasar's peculiar motion within the host in order to determine the net difference between the quasar redshift and the cloud absorption redshift. At the same time, the final position of the cloud is used to determine its fate with respect to the ionizing UV flux from the quasar.

The mean overdensity interior to the radius of the cloud $r_{\rm cloud}$ may be found from the properties of Gaussian random fields (Bardeen et al. 1986; Eisenstein & Loeb 1995). The overdensity interior to the sphere of radius $r_{\rm host} \equiv r_1$ is exactly the value of $\overline{\delta}$ that produces a collapse at the time $t_Q$. Given this constraint, we wish to find the overdensity interior to some other radius $r_{\rm cloud} \equiv r_2$. This overdensity has a Gaussian distribution with a mean and a variance that can be determined analytically. An important simplifying aspect of the problem is that we seek only to calculate the ensemble-averaged probability for observing a line at a particular redshift. This allows us to consider each radius separately instead of constructing a full realization of the spherical density profile around the host. We are free to consider the overdensity interior to $r_{\rm cloud}$ as constrained only by the existence of the host and independent of the results for other clouds. In doing so, we neglect the possibility that the mean overdensity might fail to be a monotonically decreasing function of radius, a situation that would lead to shell crossing. The effect of the latter complication on our results is expected to be rather small for the power spectra considered here. If one wished to study the correlations between cloud positions, then one would need to compute realizations of actual density profiles and compile a catalogue of absorption line spectra.

Given a power spectrum of initial density perturbations $P(k)$, the r.m.s. mass fluctuation $\delta M/M$ in a sphere of radius $r$ is

$$\sigma_M(r) = \frac{3}{\pi\sqrt{2}} \left[ \int_0^\infty dk\, P(k) \left( \frac{j_1(kr)}{r} \right)^2 \right]^{1/2}, \qquad (7)$$

where $j_1$ is the spherical Bessel function. We let $\overline{\delta}_1$ and $\overline{\delta}_2$ be the overdensities interior to radii $r_1$ and $r_2$, respectively, and then define $\nu_j \equiv \overline{\delta}_j/\sigma_M(r_j)$, where $j = 1, 2$. Given a value for $\nu_1$, the distribution of $\nu_2$ has a mean of $\gamma_{12}\nu_1$ and a variance of $1 - \gamma_{12}^2$ (Bardeen et al. 1986; Appendix A of Eisenstein & Loeb 1995), where

$$\gamma_{12} = \frac{3}{\pi\sqrt{2}\sigma_M(r_1)\sigma_M(r_2)} \int_0^\infty dk\, P(k) \left[\frac{j_1(kr_1)}{r_1}\right]\left[\frac{j_1(kr_2)}{r_2}\right]. \tag{8}$$

If the power spectrum is a power law $P(k) \propto k^n$, then one can simplify these results so that only the ratio of the mass scales enters the calculation. Manipulation of the above integrals yields the following distribution for $\overline{\delta}_2$,

$$\overline{\delta}_2 = \overline{\delta}_1 p^{-\frac{3+n}{2}} \left[g_n(p) + \xi\frac{1}{\nu_1}\sqrt{1 - g_n^2(p)}\right], \tag{9}$$

where $p = r_2/r_1$, and $\xi$ is a random number that is Gaussian distributed with zero mean and unit variance. The function

$$g_n(p) = \mathcal{N} p^{\frac{n+1}{2}} \int_0^\infty dx\, x^n j_1(x) j_1(px) \tag{10}$$

is normalized by $\mathcal{N}$ to have $g_n(1) = 1$. In this work, we use the complete (i.e. not power-law) power spectra in the Press-Schechter formalism because such detail is important for getting the correct relative weights between single galaxies and massive groups of galaxies. However, for the purpose of computing the density profiles around hosts we find it sufficient to use a simple power-law approximation to the power spectrum, which best describes the slope near the characteristic mass-scale. In particular, we use a spectral index of $n = -1.5$, which matches the index in cold dark matter cosmologies for $M \approx 10^{14}\,\mathrm{M}_\odot$.

In addition to inducing peculiar velocities, the gravitational pull of the quasar host increases the density of material and therefore the number density of Ly$\alpha$ clouds in the neighborhood of the host out to scales of order $5\,\mathrm{Mpc}$. The statistical properties of the clouds should not be disturbed in the quasi-linear regime ($\delta \lesssim 1$), as the change in their large-scale environment relative to the background universe is not significant. Under these conditions, the clouds drift smoothly together with their surrounding environment. However, as nonlinearities develop on Mpc scales the clouds may be destroyed or at least change their physical properties (e.g. neutral fraction, column density, size) due to changes in the pressure of the surrounding gas. In our analysis, we use the turn-around time ($\eta = \pi$) as the transition point between the quasi-linear and nonlinear stages ($\delta_{\mathrm{ta}} \approx 4.6$) and then assume conservatively that all clouds are destroyed after turn-around due to nonlinear hydrodynamic or thermal processes. The significance of this assumption is partially compensated by the fact that the quasar ionizes most of the Ly$\alpha$ clouds out to a distance of a few Mpc around the host cluster anyway. Thus, most of the absorbing clouds participate in quasi-linear regime flows on large scales ($\gtrsim$ Mpc), well outside the virialization shock of the host cluster or the boundary of its hot gas. It therefore seems plausible to use the above assumption as a working hypothesis. We prefer to confine our discussion to this assumption, since any of its potential modifications would add model-dependent parameters of uncertain hydrodynamics and thermodynamics to our analysis. We thus postulate that up until turn-around the line of sight to the quasar intersects more clouds as the material around the cluster drifts inwards relative to its Hubble flow. Using the spherical collapse model, the corresponding enhancement factor in the number of lines per unit radius is simply $(r_{\rm cloud}/r_H)^2$, where $r_H$ is the position that a cloud would have had in a pure Hubble flow. As a reference to the significance of our assumption, we will still show some results (cf. the uppermost curve in Figs. 1–5) for the situation where the clouds maintain their properties down to the virialization radius of their shell ($r_{\rm vir} = \frac{1}{2} r_{\rm ta}$).

The distribution of Ly$\alpha$ clouds around the quasar is diluted by the proximity effect (Carswell et al. 1982), namely the photoionization of clouds by the strong UV flux of the quasar (BDO; LWT; Kovner & Rees 1989; Bechtold 1994). The effect depends on the ratio between the UV flux of the quasar at the cloud's position and the flux of the diffuse UV background at the quasar's redshift. If the luminosity of the quasar per unit frequency at the Lyman limit is $L_\nu$, then the flux at the cloud's radius $r_{\rm cloud}$ is $F_\nu^Q = L_\nu / 4\pi r_{\rm cloud}^2$. We define $\omega = F_\nu^Q / 4\pi J_\nu$, where $J_\nu$ is the diffuse UV background flux per unit frequency at the quasar redshift at the Lyman limit. At ionization equilibrium, the neutral hydrogen column-density of each cloud is diluted by a factor $(1 + \omega)$ relative to its value in the absence of the quasar flux. Based on the power-law distribution of column densities for Ly$\alpha$ clouds, the proximity effect then implies that the number of lines above some column-density threshold is suppressed by a factor $(1 + \omega)^{-0.7}$ (BDO). To find the luminosity of the quasar, we use the observed flux $f_\nu$ at the appropriately shifted Lyman-limit frequency and multiply by $4\pi/(1 + z_Q)$ and the square of the luminosity distance in the relevant cosmology (Weinberg 1972). Because we wish to consider quasars with different fluxes as well as allow for the uncertainty in the UV background, we will do our calculations in terms of the variable

$$ f \equiv \left( \frac{f_\nu}{10^{-27}\,{\rm erg\,cm^{-2}\,s^{-1}\,Hz^{-1}}} \right) \left( \frac{10^{-21}\,{\rm erg\,cm^{-2}\,s^{-1}\,Hz^{-1}\,sr^{-1}}}{J_\nu} \right). \qquad (11) $$

Previous studies of the proximity effect have used quasars with values of $f$ in the range 1–10.

Based on the proximity effect and the two sources of peculiar velocity, we calculate the mean number of Ly$\alpha$ lines per unit velocity, $dN/dv$. The velocity difference $v$ is measured relative to the emission redshift of the quasar, so that lines with negative velocities have $z_{\rm abs} > z_Q$ and are only possible with a fortuitous coincidence of peculiar velocities. The function $dN/dv$ is obtained by integrating over the probability distributions for the quasar host masses, the quasar peculiar velocities inside the host cluster, the possible initial radii for the Ly$\alpha$ cloud, and the possible overdensities within the cloud radius. The results depend on the value of $f$, the background cosmology, and the assumed power spectrum of initial density perturbations.

## 3. Results

We study two cosmologies, each of which gives a rough fit to the observed large scale structure in the local universe. They both involve cold dark matter (CDM) with a power spectrum of the form

$$P(k) \propto \frac{k}{\left\{1 + [ak + (bk)^{3/2} + (ck)^2]^\nu\right\}^{2/\nu}} \qquad (12)$$

where $a = (6.4/\Gamma)h^{-1}$ Mpc, $b = (3.0/\Gamma)h^{-1}$ Mpc, $c = (1.7/\Gamma)h^{-1}$ Mpc, and $\nu = 1.13$ (Efstathiou, Bond, & White 1992). For our flat cosmology, we use $\Omega = 1$, $\Gamma = 0.5$, and $\sigma_{8h^{-1}\,\mathrm{Mpc}} = 0.6$. For our open cosmology, we use $\Omega = 0.3$, $\Gamma = 0.2$, $\sigma_{8h^{-1}\,\mathrm{Mpc}} = 0.8$, and a zero cosmological constant. The value of the Hubble constant cancels out of the calculation, except for the indirect dependence in $\Gamma$. All masses in this paper refer to the total mass, including dark matter.

We present results for two different redshifts, $z = 2$ and $z = 4$, and for various values of $f$. We consider three different cases. The first is referred to as *"Proximity Only"*, where we set all the peculiar velocities to zero and include only the proximity suppression $(1+\omega)^{-0.7}$, with $r_{\mathrm{cloud}}$ inferred from the Hubble flow. This case has been used in previous analyses of the proximity effect. It describes a situation where quasars preferentially reside in low mass systems with negligible peculiar velocities. The second case is referred to as *"Infall, $M > 10^{12}\,M_\odot$"* and includes all the sources of peculiar velocity that were discussed in §2. In this case, we assume that the minimum mass of the quasar host is $10^{12}\,M_\odot$. The third case is referred to as *"Infall, $M > 10^{13}\,M_\odot$"* and is the same as the second case except that the minimum mass of the host is $10^{13}\,M_\odot$. This means that quasars are not allowed to exist in field galaxies and instead reside in systems at least as massive as groups of galaxies. This condition increases the velocity dispersion that the quasar can have within the host system and extends the gravitational effect of the host on the Ly$\alpha$ clouds. The third case is suggested by the fact that the correlation function of quasars is closer to that of groups than to that of field galaxies (Bahcall & Chokshi 1991).

In Figures 1–4, we present plots of the expected number of lines per unit velocity $dN/dv$. Positive velocities mean $z_{\mathrm{abs}} < z_Q$. The density of lines $dN/dz$ far away from the quasar was assumed to be 350 for $z = 2$ and 700 for $z = 4$ (see discussion following Eq. 6); all predictions in these figures and in the tables are linearly proportional to these values. Each figure represents a different combination of redshift and cosmology. Within the figures, each set of three curves shows the results for three values of $f$. The case $f = 0$ means that the proximity effect has been turned off; the *"Proximity Only"* case then represents the base number of lines that would be expected far from the quasar. As $f$ gets larger, the quasar is brighter and there are fewer lines near the quasar due to the increase in its ionizing flux. In addition to the above cases, we also show with the dot-dash line the extreme case of infall with $M > 10^{12} \, M_\odot$ and $f = 0$ in which we allow the clouds to proceed past turn-around all the way to the virial radius of their shell ($\eta = 3\pi/2$). Note that while this case produces a marked increase for $v < 1000 \, \mathrm{km/s}$, it makes no change relative to the *"Infall, $M > 10^{12} \, M_\odot$"* case for $v > 1000 \, \mathrm{km/s}$.

In light of recent spectroscopic observations with the Hubble Space Telescope (Bahcall et al. 1993), it is interesting to consider the clustering signature at low redshifts. The shapes of the $dN/dv$ curves at $z < 1$ are similar to those presented in Figures 1 and 2, but because of the scarcity of Ly$\alpha$ lines at low redshifts, the normalization of the vertical axis is very different. Not only does the abundance of clouds fall steeply at low redshift, but also the equivalent width threshold of HST is much greater than that of the Keck telescope. For HST observations at $z \approx 1$, we estimate $dN/dz < 20$, which means that the number of lines with $v < 0 \, \mathrm{km/s}$ is undetectably small. In addition, the background flux $J_\nu$ estimated by Kulkarni & Fall (1994) at $z < 1$ is smaller by 1–2 orders of magnitude than the estimated background at $2 < z < 4$. This means that for a given intrinsic quasar luminosity, the proximity effect extends out to a much larger radius of influence at $z < 1$, further concealing the effects of clustering. Put another way, Figures 1–4 show that the number of lines with negative velocity shifts is significantly smaller than the deficit caused by the proximity effect. Since the detection of the proximity effect at low redshift (Kulkarni & Fall 1994) is only marginal, the predicted clustering signature is most likely undetectable with HST for $z < 1$.

In general, the inclusion of infall and peculiar velocities increases the number of lines relative to the proximity effect. This is primarily caused by the increase in the number density of clouds around the host. For bright quasars, the higher ionizing flux extends the proximity effect to a large radius. The surviving Ly$\alpha$ clouds only feel a weak gravitational pull; moreover, the peculiar velocity of the quasar within its host is only a small fraction of the Hubble velocity of the clouds. Thus, the distribution of the clouds is only weakly perturbed from that of the proximity effect alone. Dim quasars, however, cannot ionize the clouds to a sufficiently large radius so as to hide the effects of infall and peculiar velocities. These quasars develop a tail of negative velocity lines as well as a significant excess of lines for $v \lesssim 2000 \, \mathrm{km/s}$. Note that in the limit of $f \to 0$, the number of Ly$\alpha$ lines per unit redshift is not conserved because the infall allows a gathering of clouds from the two dimensions transverse to the line of sight; this is in difference from the effect of the quasar motion inside the host, which merely shifts the velocity distribution of lines without changing their total number. As revealed by the excess bump in the number of Ly$\alpha$ lines shown in Figures 1–4 for $f = 0$, it is even likely to find an "anti-proximity" effect near very dim quasars.

We expect that a similar statistical excess of Ly$\alpha$ lines will occur near galaxies along the line of sight, such as the ones associated with metal absorption systems (Barcons & Webb 1990). In these cases, the proximity effect of the galaxies should be much weaker than that of quasars and the extent of the excess bump should be closely described by the $f = 0$ curves. We investigate this by assuming that the metal-line systems are metal-rich clouds in the halos of galaxies (Steidel & Sargent 1992). We neglect the impact parameter between the line of sight and the center of the host and then allow the metal-line system itself to move with the velocity dispersion of the host halo. The expected $dN/dv$ distribution of Ly$\alpha$ lines relative to the host is then found by taking the $dN/dv$ curve for the quasar case (Fig. 1-4) and adding it to its reflection around $v = 0$. The result is shown in Figure 5. Here we consider both $M > 10^{11}\,M_\odot$ and $M > 10^{12}\,M_\odot$ for $\Omega = 0.3$, as well as $M > 10^{12}\,M_\odot$ for $\Omega = 1$. We also show the result for $M > 10^{12}\,M_\odot$ and $\Omega = 0.3$ if one allows the clouds to evolve beyond turn-around and fall back to the virial radius of their shell. In all cases, one finds an appreciable increase in the numbers of Ly$\alpha$ clouds around metal-line systems. Integrating the difference between the curves in Figure 5 and the base expectations (dotted line) yields an excess of 1.5 Ly$\alpha$ lines per system in the $M > 10^{11}\,M_\odot$ case, 2.1 lines in the $M > 10^{12}\,M_\odot$, $\Omega = 0.3$ case, 0.9 lines in the $M > 10^{12}\,M_\odot$, $\Omega = 1$ case, and 8.6 lines in the *"Beyond Turn-Around"* case. As before, these numbers are linearly proportional to $[(dN/dz)/350]$.

This effect has been searched for using the cross-correlation of metal-line systems with Ly$\alpha$ lines (Barcons & Webb 1990), but no significant signal was found. The number of excess lines predicted by our model for this study may be estimated as follows. The study included 51 metal-line systems, many around $z = 2$ but a few at higher redshift. Adopting $z = 2$, the equivalent width threshold of 0.36Å yields $dN/dz = 40$ (BDO). If $1.5 \times [(dN/dz)/350]$ excess lines are expected around each metal-line system, then around 9 extra lines would be present in the sample, presumably a detectable number. However, the Ly$\alpha$ line that is associated with the metal-line system itself (which, of course, is not included in the cross-correlation) tends to be broad and therefore obscures other lines within several hundred km/s of the associated line (Barcons & Webb 1990). As can be seen from Figure 5, removing the range $-300\,\mathrm{km/s} < v < 300\,\mathrm{km/s}$ cuts the signal by about a factor of two. If one uses $\Omega = 1$ rather than $\Omega = 0.3$, then the signal is further reduced by another factor of two, yielding an excess of 2–3 lines, which is marginally compatible with the null result of the cross-correlation study.

The more extreme *"Beyond Turn-Around"* model is ruled out by these observations. Studies with higher resolution and lower equivalent width threshold should strongly constrain these models; continued null results would require lower masses for the metal-line systems, or the destruction of the Ly$\alpha$ clouds at a lower density threshold.

The above results for $dN/dv$ are ensemble averages in which we have integrated over several unknown parameters, such as the mass of the host. This means that for a particular quasar, the positions of the absorption lines will be correlated. For example, if the quasar resides in a massive cluster this would favor having multiple low velocity lines, but if the quasar resides in a field galaxy, then the chance of having low velocity lines is small. Therefore, if one line is seen with $v < 0$, the chance of seeing another low velocity line is increased.

Tables 1 and 2 present the expected number of lines per quasar with negative velocites. Table 1 is for $z = 2$, and Table 2 is for $z = 4$. For each statistic, we show results in the two cosmologies. The $\Omega = 1$ cosmology, with its more recent structure formation, produces a slightly smaller tail at $v < 0$. Because we remove any clouds that evolve past turn-around relative to the quasar host, there would be no negative velocity lines were it not for the peculiar velocity of the quasar within its host. Relaxing the condition that no absorption clouds survive beyond turn-around and instead allowing them to survive down to the virial radius of the shell produces a stronger negative velocity tail. For bright quasars, the number of lines with $v < 0$ are doubled and the number with $v < -500$ km/s are tripled; for dimmer quasars, the increase is even larger. The increase is not only caused by the fact that the clouds beyond turn-around have a higher redshift than the quasar host but also by the large density enhancements ($\sim$180 above background) that are reached as the shell containing the clouds approaches virialization.

Table 3 presents the total number of lines expected between $v = 0$ and $v = 3000$ km/s for $z = 2$ and a variety of $f$ values. The numbers listed are the difference between the calculated number and the base number that would be expected within the same velocity range far away from the quasar. Table 4 presents the results for $z = 4$. The above velocity range was picked in order to roughly maximize the signal of the deficit of lines compared to the noise of the base level. In principle, the signal-to-noise ratio can be further improved by picking different ranges for different values of $f$ and $z$.

Tables 1–4 show minor differences between the two CDM cosmologies. The results for $\Omega = 1$ show less infall and smaller peculiar velocities, since in an $\Omega = 1$ universe structure forms later than in an open universe (Richstone, Loeb, & Turner 1992). We also tested the sensitivity of the results to changes in the assumed spectral index $n = -1.5$ around the host mass-scale by trying values of $n = -2$ and $n = -1$. As expected, $n = -2$ yields a stronger infall signal, while $n = -1$ results in a weaker one. For the production of lines with $v < 0$, the differences are less than 20%. For the number of lines between 0 and $3000\,\mathrm{km/s}$, the $n = -2$ spectrum produces a significantly larger excess because it makes the density fluctuations much larger at radii of order $5\,\mathrm{Mpc}$.

Detection of a deficit of lines below the base expectations has been taken as evidence for the proximity effect and then used to determine the UV background flux (BDO, LWT, Bechtold 1994). The addition of peculiar velocities changes the predicted deficit of lines, although for large values of $f$ the difference is small. Therefore, one should use bright quasars (with $f \gtrsim 1$) to determine the UV background flux. For dim quasars ($f \lesssim 1$), neglecting peculiar velocities would lead to a significant overestimate of the UV background. To get the size of this effect, one should use the number in Table 3 or 4 for a particular value of $f$ in one of the *"Infall"* cases and then find the value of $f$ in the *"Proximity Only"* case that gives the same deficit. The ratio of the two $f$ values estimates the error in the inferred UV background. Applying this procedure to Table 3 suggests that the standard approach to the proximity effect overestimates the background flux by up to a factor of $\sim 3$. For example, a deficit of about 1.6 lines would indicate $f = 1$ with infall but only $f \approx 0.4$ without infall, hence the UV background flux would have been overestimated by a factor of 2.5 if infall was ignored. In making this estimate, we have only considered lines with $v > 0$ because candidate Ly$\alpha$ lines with $v \lesssim 0$ have typically been marked as unidentified and omitted. In addition, the uncertainties in the quasar redshift are another source of error in the inferred UV background (Espey 1993; Bechtold 1994). The effect of infall would be particularly pronounced in samples of radio-loud quasars, which are known to reside in rich cluster environments (Yee & Green 1987). This clustering signature has important implications as it tends to lower the inferred background flux down to a level that could be provided by the known populations of quasars (Miralda-Escudé & Ostriker 1990; Madau 1992; Fall & Pei 1993; Meiskin & Madau 1993).

While more massive hosts are more effective at producing peculiar velocity shear, the small probability that the quasars actually reside in such hosts keeps these mass scales from dominating the results. For $f = 3$ at $z = 2$, the contribution to the number of lines with $v < 0$ for hosts of $10^{11}\,\mathrm{M_\odot}$, $10^{12}\,\mathrm{M_\odot}$, $10^{13}\,\mathrm{M_\odot}$, and $10^{14}\,\mathrm{M_\odot}$ *per host site* per $\log M$ are in the ratio $1.0 : 3.4 : 13.5 : 53.2$. This indicates that for brighter quasars the high mass systems are much more efficient at producing negative velocity lines. However, when we weight these contributions by the actual numbers of such hosts for $\Omega = 0.3$, we get the ratio $1.0 : 3.6 : 8.3 : 1.9$. Systems with masses between $10^{12}\,\mathrm{M_\odot}$ and $6 \times 10^{13}\,\mathrm{M_\odot}$ provide most of the signal. Indeed, the fact that there is a difference between the $10^{12}\,\mathrm{M_\odot}$ and $10^{13}\,\mathrm{M_\odot}$ minimum mass cases demonstrates that the massive $\gtrsim 10^{14}\,\mathrm{M_\odot}$ clusters are not dominating the results. If one assumes that quasars reside in lower mass ($M \approx 10^{11}\,\mathrm{M_\odot}$) hosts, then the small mass of these hosts will produce little deviation from the usual proximity effect. In a similar fashion, any inaccuracies in the Press-Schechter approximation that yield the wrong ratio between the number density of field galaxies and groups of galaxies will alter the results roughly linearly in the ratio between the number densities of these two mass scales. It is worth noting that for $f = 0$, even small mass scales are capable of producing an appreciable number of lines near $v = 0$, but these lines are the first to disappear when the proximity effect is turned on.

A related concern is that at high redshift the number density of quasars may exceed the number density of objects with masses greater than $10^{13}\,M_\odot$. At $z = 2$, we find the comoving number density of objects with masses greater than $10^{13}\,M_\odot$ to be $9\times10^{-5}h^3\,\mathrm{Mpc}^{-3}$ in the open cosmology and $8 \times 10^{-5}h^3\,\mathrm{Mpc}^{-3}$ in the flat cosmology. At $z = 4$, we find $8 \times 10^{-6}h^3\,\mathrm{Mpc}^{-3}$ in the open cosmology and $1.5 \times 10^{-7}h^3\,\mathrm{Mpc}^{-3}$ in the flat cosmology. Since the comoving number density of quasars at the relevant magnitude range is $\sim 10^{-6}h^3\,\mathrm{Mpc}^{-3}$ (Hartwick & Schade 1990; Schmidt, Schneider, & Gunn 1991), these results suggest that quasars at $z = 4$ in a flat cosmology must have a minimum host mass less than $10^{13}\,M_\odot$. Furthermore, to make the $\Omega = 1$ results in the tables match those of the $\Omega = 0.3$ cosmology at a given value of $f$, one has to increase the minimum host mass, thus pushing against the limited number density of massive hosts in the flat universe. When we apply our model to mixed dark matter (MDM) cosmologies (Klypin et al. 1993; Klypin et al. 1994; Mo & Miralda-Escudè 1994; Ma & Bertschinger 1994), we run out of hosts with masses greater than $10^{12}\,M_\odot$ as we approach $z = 4$. One is therefore forced to consider lower mass hosts, which strongly reduces the infall effects relative to the proximity effect. Therefore, models like MDM that form structure late should be well described by the *"Proximity Only"* case at high $z$ and will produce few, if any, negative velocity lines.

## 4. Conclusions

In this paper we have quantified the signature of early clustering on the redshift distribution of Ly$\alpha$ lines around the quasar redshift. We have shown that the inclusion of peculiar velocities and density enhancements in the neighborhood of quasar hosts leads to an excess of lines for redshifts within $0.01(1 + z_Q)$ of the quasar redshift (i.e. $v < 3000\,\mathrm{km/s}$) and allows some lines to have redshifts higher than the quasar. The effect is primarily caused by the peculiar velocity shear and the corresponding overdensities on large scales ($\gtrsim$ Mpc) and is dominated by the contributions of groups of galaxies rather than by rich clusters. In the linear regime of structure formation ($\delta\rho/\rho \ll 1$), the environment of the clouds does not change significantly relative to the smooth background. In our calculations, we assumed that the statistical properties of Ly$\alpha$ clouds are not altered even in the quasi-linear regime as they drift smoothly together with their large-scale environment, but that the clouds are entirely destroyed in the nonlinear infall phase. The transition point between the quasi-linear and nonlinear stages was primarily assumed to be turn-around relative to the quasar host. The actual fate of Ly$\alpha$ clouds in the nonlinear infall stages depends on the physical mechanism responsible for the confinement of the clouds, which is at present uncertain (Cen et al. 1994, and references therein).

The statistics of lines with negative velocity shifts relative to the quasar provides a measure of structure at high redshifts ($1 \lesssim z \lesssim 5$). If quasars typically reside in small groups of galaxies ($\sim 10^{13} M_\odot$), then the clustering signature should already show-up in samples of a few quasars observed with the Keck telescope (cf. Tables 1-4). The association of quasars with groups is suggested by their own clustering properties (Bahcall & Chokshi 1991) and their cross-correlation with galaxies (Hartwick & Schade 1990; Ellingson, Green, & Yee 1991). A hint for unidentified absorption lines, which could in principle be Ly$\alpha$ with $z_{\rm abs} \gtrsim z_Q$, is already found in existing data sets (Bechtold 1994). Weaker effects could be detected with larger quasar samples, such as provided by the Sloan Digital Sky Survey (Gunn & Knapp 1993). The clustering signal should appear more frequently in samples of radio-loud quasars, since these objects are known to reside in richer cluster environments than radio-quiet quasars (Yee & Green 1987). Indeed, an excess of associated C IV absorbers was found in a sample of radio-loud quasars (Anderson et al. 1987) and was not seen for radio-quiet objects (Young et al. 1982).

The main practical obstacle towards measuring the clustering signature is the observational uncertainty in quasar redshifts. The significance of the uncertainty is emphasized by the steep exponential decline in the number of lines at negative velocity shifts (cf. Figs. 1-4); small shifts in the velocity threshold can cause a significant change in the expected number of lines. A quasar redshift cannot be determined accurately from its Ly$\alpha$ emission line, since the blue side of the line is distorted by the Ly$\alpha$ forest. However, the C IV emission line can set the quasar redshift to within $\pm 300\,{\rm km/s}$ (Tytler & Fan 1992; Laor et al. 1994). Infrared observations with high spectral resolution of the [O III] narrow emission line can further improve the redshift determination of the quasar to better than $\pm 50\,{\rm km/s}$. Narrow lines are thought to originate from the host galaxy and therefore provide the relevant redshift for our analysis (Vrtilek & Carleton 1985).

The identification and interpretation of Ly$\alpha$ absorption lines beyond the quasar redshift could be complicated by various factors. First, such lines may be explained as metal absorption lines. However, based on the background frequency of metal absorption lines, the probability that a metal line from a different redshift will randomly appear within $1000\,{\rm km/s}$ of the Ly$\alpha$ emission line of the quasar is very low. In addition, this potential confusion can be resolved by tracing the Ly$\beta$ counterparts of the Ly$\alpha$ lines. Second, a negative velocity line may be interpreted as being unrelated to cosmological peculiar velocities but rather as being caused by Ly$\alpha$ absorption from a cloud inside the host galaxy. If an absorption cloud has a large relative velocity due to the fact that it resides within the broad line region of the quasar, then it is likely to have a metalicity higher than solar (Hamann & Ferland 1992) and its Ly$\alpha$ redshift should show-up also in metal absorption lines. If associated metal lines are not found, the remaining possibility is that the cloud resides in a metal poor environment bound within the host galaxy, in which case the velocity shifts are limited to $\lesssim 300$ km s$^{-1}$ and can be distinguished from cluster velocities.

The Ly$\alpha$ clouds have a characteristic response time of $\sim 10^4$ yr/$(1+\omega)$ to variations in the quasar flux. If quasars vary considerably on this timescale, then the observed redshift distribution of Ly$\alpha$ lines may correspond to a different value of $f$ than the apparent quasar fluxes (BDO). While this may complicate somewhat the conversion of $dN/dv$ to constraints on cosmological models, it will still allow the clustering signature to appear in all cases where the proximity effect is only secondary to the redshift distortion caused by clustering (cf. Tables 3 and 4).

Clustering at high redshifts has other implications aside from the existence of lines with negative velocity shifts. First, the redshift distribution of Ly$\alpha$ lines should show small non-zero correlations along any random line of sight due to the existence of mass concentrations (Sargent et al. 1980; Webb & Barcons 1991). The detection of these correlations may prove to be more difficult than finding Ly$\alpha$ lines beyond the quasar redshift since quasars preferentially reside in dense environments. However, near the redshift of very dim quasars or metal absorption lines, a significant statistical excess of $\gtrsim 1.5 \times [(dN/dz)/350]$ Ly$\alpha$ lines is expected at $z = 2$ (cf. Fig. 5). Existing samples (Barcons & Webb 1990) are already at the threshold of testing this signature, while higher resolution and more sensitive data available with the Keck telescope should easily reveal this excess. A second implication of clustering is that the standard approach to the proximity effect overestimates the ionizing background flux at high redshifts, especially in samples of faint radio-loud quasars. For published samples (BDO, LWT, Bechtold 1994), the clustering effect could lower the inferred background flux by up to a factor of $\sim 3$ (cf. Tables 3 and 4). This tends to weaken the discrepancy between the deduced background flux and the contribution from known populations of quasars (see also Espey 1993; Bechtold 1994). It is possible to reduce the clustering effect on the estimation of the proximity effect by considering intrinsically bright quasars; the values of the three cases in Tables 3 and 4 converge as $f$ increases.

Finally, we emphasize that the numbers in this paper were derived for specific (CDM) cosmological models and would be significantly lower in models with later structure formation

(e.g. MDM). The strength of a detected clustering signature can be used to infer a lower limit on the number density of groups of galaxies at high redshifts and to discriminate among these models.

We thank J. Bahcall, A. Dobrzycki, A. Laor, E. Maoz, W. Sargent, C. Steidel, J. Webb, and especially J. Ostriker for useful discussions. D.J.E. was supported in part by a National Science Foundation Graduate Research Fellowship.

| Case | | $f = 0$ | $f = 0.1$ | $f = 0.3$ | $f = 1$ | $f = 3$ | $f = 10$ |
|---|---|---|---|---|---|---|---|
| $v < -500\,{\rm km/s}$ | $\Omega = 0.3$ | 0.050 | 0.044 | 0.037 | 0.026 | 0.016 | 0.008 |
| $M > 10^{12}\,{\rm M}_\odot$ | $\Omega = 1.0$ | 0.027 | 0.022 | 0.017 | 0.011 | 0.006 | 0.003 |
| $v < 0\,{\rm km/s}$ | $\Omega = 0.3$ | 0.56 | 0.45 | 0.35 | 0.23 | 0.13 | 0.064 |
| $M > 10^{12}\,{\rm M}_\odot$ | $\Omega = 1.0$ | 0.45 | 0.35 | 0.26 | 0.16 | 0.088 | 0.041 |
| $v < 0\,{\rm km/s}$ | $\Omega = 0.3$ | 0.88 | 0.79 | 0.67 | 0.48 | 0.30 | 0.15 |
| $M > 10^{13}\,{\rm M}_\odot$ | $\Omega = 1.0$ | 0.77 | 0.69 | 0.58 | 0.40 | 0.24 | 0.12 |

Table 1: The expected number of lines per quasar with a sufficiently negative velocity. The redshift of the quasar is 2. The numbers are proportional to $[(dN/dz)/350]$.

| Case | | $f = 0$ | $f = 0.1$ | $f = 0.3$ | $f = 1$ | $f = 3$ | $f = 10$ |
|---|---|---|---|---|---|---|---|
| $v < -500\,{\rm km/s}$ | $\Omega = 0.3$ | 0.11 | 0.040 | 0.022 | 0.010 | 0.005 | 0.002 |
| $M > 10^{12}\,{\rm M}_\odot$ | $\Omega = 1.0$ | 0.077 | 0.021 | 0.011 | 0.005 | 0.002 | 0.001 |
| $v < 0\,{\rm km/s}$ | $\Omega = 0.3$ | 1.71 | 0.53 | 0.28 | 0.13 | 0.062 | 0.027 |
| $M > 10^{12}\,{\rm M}_\odot$ | $\Omega = 1.0$ | 1.55 | 0.44 | 0.23 | 0.10 | 0.049 | 0.021 |
| $v < 0\,{\rm km/s}$ | $\Omega = 0.3$ | 3.00 | 1.48 | 0.88 | 0.43 | 0.21 | 0.091 |
| $M > 10^{13}\,{\rm M}_\odot$ | $\Omega = 1.0$ | 2.85 | 1.46 | 0.87 | 0.43 | 0.21 | 0.091 |

Table 2: The expected number of lines per quasar with a sufficiently negative velocity. The redshift of the quasar is 4. The numbers are proportional to $[(dN/dz)/700]$.

| Case | | $f = 0$ | $f = 0.1$ | $f = 0.3$ | $f = 1$ | $f = 3$ | $f = 10$ |
|---|---|---|---|---|---|---|---|
| Proximity | $\Omega = 0.3$ | 0.0 | −0.83 | −1.39 | −2.40 | −3.79 | −5.76 |
| Only | $\Omega = 1.0$ | 0.0 | −0.86 | −1.45 | −2.49 | −3.92 | −5.91 |
| Infall | $\Omega = 0.3$ | 0.40 | −0.01 | −0.53 | −1.60 | −3.12 | −5.28 |
| $M > 10^{12}\,M_\odot$ | $\Omega = 1.0$ | −0.02 | −0.43 | −0.96 | −2.01 | −3.51 | −5.60 |
| Infall | $\Omega = 0.3$ | 0.52 | 0.23 | −0.21 | −1.21 | −2.73 | −4.96 |
| $M > 10^{13}\,M_\odot$ | $\Omega = 1.0$ | −0.07 | −0.34 | −0.77 | −1.74 | −3.21 | −5.35 |

Table 3: The change in the expected number of lines with $0 < v < 3000\,\mathrm{km/s}$ from the number expected far from the quasar: 10.5. The quasar is at redshift 2. The numbers are proportional to $[(dN/dz)/350]$.

| Case | | $f = 0$ | $f = 0.1$ | $f = 0.3$ | $f = 1$ | $f = 3$ | $f = 10$ |
|---|---|---|---|---|---|---|---|
| Proximity | $\Omega = 0.3$ | 0.0 | −9.4 | −14.6 | −21.4 | −27.1 | −31.2 |
| Only | $\Omega = 1.0$ | 0.0 | −9.4 | −14.5 | −21.3 | −26.5 | −31.2 |
| Infall | $\Omega = 0.3$ | 0.58 | −7.4 | −13.0 | −20.3 | −26.3 | −30.8 |
| $M > 10^{12}\,M_\odot$ | $\Omega = 1.0$ | −0.40 | −8.0 | −13.4 | −20.5 | −26.4 | −30.8 |
| Infall | $\Omega = 0.3$ | 0.80 | −6.0 | −11.5 | −19.1 | −25.6 | −30.4 |
| $M > 10^{13}\,M_\odot$ | $\Omega = 1.0$ | −0.71 | −6.8 | −12.1 | −19.4 | −25.7 | −30.4 |

Table 4: The change in the expected number of lines with $0 < v < 3000\,\mathrm{km/s}$ from the number expected far from the quasar: 35.0. The quasar is at redshift 4. The numbers are proportional to $[(dN/dz)/700]$.

## FIGURE CAPTIONS

**Figure 1:** The number density of Ly$\alpha$ lines per unit velocity in units of $(10^3\,\mathrm{km/s})^{-1}$ for the $\Omega = 0.3$ cosmology at $z = 2$. The results are shown for three values of $f$, each with curves for the two *"Infall"* cases and for the *"Proximity Only"* case. Negative velocities mean $z_{\mathrm{abs}} > z_Q$. The dot-dashed curve shows the $f = 0$, $M > 10^{12}\,\mathrm{M}_\odot$ result for the case where the Ly$\alpha$ clouds are not destroyed until they reach their virialization radius. The numbers on the vertical axis are proportional to $[(dN/dz)/350]$.

**Figure 2:** The same as figure 1, but for $\Omega = 1$ and $z = 2$. The numbers on the vertical axis are proportional to $[(dN/dz)/350]$.

**Figure 3:** The same as figure 1, but for $\Omega = 0.3$ and $z = 4$. The numbers on the vertical axis are proportional to $[(dN/dz)/700]$.

**Figure 4:** The same as figure 1, but for $\Omega = 1$ and $z = 4$. The numbers on the vertical axis are proportional to $[(dN/dz)/700]$.

**Figure 5:** The number density of Ly$\alpha$ lines per unit velocity in units of $(10^3\,\mathrm{km/s})^{-1}$ around a metal-line system at $z = 2$. The solid and dot-dashed curves are for the *"Infall"* model with $f = 0$ and minimum masses of the host of the metal-line system of $10^{12}\,\mathrm{M}_\odot$ for $\Omega = 0.3$ and $\Omega = 1$, respectively. The short-dashed curves is for the *"Infall"* model with a minimum host mass of $10^{11}\,\mathrm{M}_\odot$ and $\Omega = 0.3$. The long-dashed curve assumes $M > 10^{12}\,\mathrm{M}_\odot$ and $\Omega = 0.3$ and allows the clouds to survive past turn-around and until virialization. The dotted curve is the base number density of clouds in the absence of any clustering. The numbers on the vertical axis are proportional to $[(dN/dz)/350]$.